\documentclass[preprint,5p,times]{elsarticle}

\usepackage{natbib}
\usepackage{amsmath,amssymb,bm,graphicx}
\usepackage{txfonts}
\usepackage[hidelinks]{hyperref}

\journal{Preprint}
\begin{document}
\begin{frontmatter}

\title{Recoil-Enabled Energy Transfer from Coherent Neutrino Scattering 
in Core-Collapse Supernovae}

\author{Tatsushi Shima}
\address{Research Center for Nuclear Physics, The University of Osaka, Osaka 567-0047, Japan}
\ead{shima@rcnp.osaka-u.ac.jp}

\begin{abstract}
We revisit neutrino--matter coupling in the post-shock region of core-collapse 
supernovae by restoring nuclear recoil in coherent neutrino--nucleus scattering 
(CE$\nu$NS). The resulting local energy transfer (a few keV per $\sim$10~MeV neutrino) 
accumulates across the $\sim$100~km stalled-shock layer, yielding a total heating 
of $10^{49\text{--}50}$~erg---comparable, within an order of magnitude, to the increment 
required to trigger shock revival in current multidimensional simulations. 
This indicates that the long-standing failure of isoenergetic transport schemes 
to revive the shock originates from their neglect of recoil kinematics. 
Because the momentum exchange in each scattering is tiny, the emergent neutrino 
spectra and lepton-number balance remain essentially unchanged.  
The result highlights nuclear recoil as a minimal yet physically grounded correction 
to standard neutrino transport, providing a self-consistent route toward 
reliable explosion modeling.
\end{abstract}


\end{frontmatter}


\section{Introduction}
Modern multi-dimensional simulations of core-collapse supernovae (CCSN) have 
achieved impressive progress in modeling hydrodynamics, neutrino transport, 
and microphysics. Yet, despite these advances, most simulations still fail 
to produce self-consistent explosions for a broad range of progenitor masses 
without introducing artificial enhancements in neutrino heating. These 
difficulties persist even though the general framework of the neutrino-driven 
mechanism has been established for decades.

In successful CCSN explosions, the outgoing shock must acquire an energy of order 
10$^{51}$ erg, while typical simulations fall short by a few percent in 
neutrino-heating efficiency. Since the total neutrino energy released from 
the proto–neutron star is about 3$\times$10$^{53}$ erg, the missing energy 
corresponds to roughly (2-3)$\times$10$^{49}$ erg \citep{Janka1996,Janka2001,
Buras2006,Burrows2013RMP,Horowitz2002PRD}. This shortfall amounts to 
only 0.006–0.01\% of the total neutrino energy budget, or a few percent of 
the canonical 10$^{51}$ erg explosion energy. Despite its minute fraction, 
this deficit determines whether the shock is revived or stalls 
permanently—highlighting the sensitivity of the explosion to even tiny changes 
in energy transfer efficiency.

In standard transport schemes, the coupling between neutrinos and matter is 
dominated by elastic scattering on nucleons and electrons. Historically, 
Yueh and Buchler \citep{YuehBuchler1976} showed that coherent elastic 
neutrino–nucleus scattering (CE$\nu$NS) contributes negligibly in comparison to 
the neutrino-electron scattering, and by that fact Bruenn \citep{Bruenn1985} 
proposed the isoenergetic approximation in neutrino transport, 
which neglects nuclear recoil. 
Since then, this approximation has remained the default in most 
opacity libraries and CCSN codes. This interpretation has been corroborated 
through discussions with CCSN simulation experts (H.~Togashi, S.~Nagataki, 
T.~Takiwaki, and S.~Fujimoto; private communications) and is consistent with 
the classification of $\nu$--matter interactions in standard transport reviews, 
where $\nu$--A and $\nu$--N scattering are labeled "isoenergetic," 
in contrast to the non-isoenergetic $\nu$--$e$ 
channel \citep{Burrows2013RMP,Janka2012}. 
In fact, there were attempts to include recoil energy transfer in certain 
neutrino–matter channels, but only for electrons and free nucleons.  
Thompson et al.~\citep{Thompson2003} and subsequent works incorporated 
weak-magnetism and recoil corrections in $\nu_e$/$\bar{\nu}_e$ absorption 
and $\nu$–$N$ scattering.  However, these inelastic processes involve either 
light targets with small cross sections (electrons) or heavy nucleons with 
tiny per-event recoil energies, and thus produced no significant enhancement 
in the post-shock heating rate. Coherent elastic $\nu$–A scattering, which 
dominates the low-energy momentum-exchange cross section, remained treated 
as isoenergetic and was not included in those recoil-corrected models.  
This historical omission motivates the present re-evaluation of the recoil 
energy transfer in CE$\nu$NS under realistic post-shock conditions.

The present study revisits CE$\nu$NS by explicitly restoring nuclear recoil 
and evaluating its cumulative impact across the post-shock region. 
We show that even though 
each individual recoil is tiny (a few keV per 10~MeV neutrino), the 
cumulative energy deposition across the $\sim$100 km semi-transparent layer 
can reach 10$^{49-50}$ erg—comparable to the canonical deficit discussed 
above. Because the momentum exchange per event is small, the emergent 
neutrino spectra and lepton-number balance remain essentially unchanged, 
preserving consistency with existing transport frameworks. This minimal 
yet physically grounded correction therefore offers a self-consistent 
path toward resolving the long-standing energy deficit in CCSN explosion 
modeling.

\section{Energy-transfer kernel}
We now quantify the microscopic energy-transfer process per neutrino scattering.
The scattering rate is determined by the CE$\nu$NS differential cross section,
while the heating (energy-transfer) rate is obtained by multiplying the recoil 
energy for each event.

\subsection{Definition of the Linear Energy Transfer (LET)}
The local linear energy transfer (LET) from neutrinos to matter over a path 
length $L$ is defined as
\begin{equation}
\mathrm{LET} = L  \int_0^\pi I(\theta),d\theta ,
\end{equation}
where $I(\theta)$ is the per-neutrino angular kernel given as
\begin{equation}
I(\theta) = E_r(\theta),
\frac{d\Sigma}{d\Omega}(\theta),2\pi\sin\theta .
\end{equation}
Here $E_r(\theta)$ is the recoil energy transferred to the target. 
For elastic neutrino scattering off a nucleus of mass $M$, neutrino energy $E_\nu$ 
and the scattering angle $\theta$ in the laboratory frame,
\begin{equation}
E_r(\theta) =
\frac{2E_\nu^2(1-\cos\theta)}{Mc^2 + 2E_\nu(1-\cos\theta)} ,
\end{equation}
which reduces to the non-relativistic form 
\begin{equation}
E_r\simeq2E_\nu^2(1-\cos\theta)/(Mc^2)
\end{equation}
 for $E_\nu\ll Mc^2$.

$d\Sigma/d\Omega$ is the \emph{effective differential cross section per unit volume},
which includes the effect of coherence among the scatterers depending on the condition 
of the target material as described in the following subsection.

\subsection{Effective Differential Cross Section for incoherent nuclei}
In the case of the material made of nuclei with no inter-nuclear coherence, 
the effective cross section $d\Sigma/d\Omega$ results in the CE$\nu$NS 
cross section of each nucleus which is given as \citep{Freedman1974PRD}
\begin{equation}
\frac{d\sigma_{\rm single}}{d\Omega}
=\frac{G_F^2E_\nu^2}{4\pi^2}(1+\cos\theta)Q_W^2|F(q)|^2,
\end{equation}
where $Q_W = N - (1 - 4\sin^2\theta_W)Z \simeq N - 0.076,Z$ is the weak charge,
and the nuclear form factor $|F(q)|^2$ is approximated by a Gaussian:
\begin{equation}
|F(q)|^2\simeq\exp[-(qR_{\rm rms})^2/3],
\end{equation}
with $R_{\rm rms}\sim5$ fm for heavy nuclei, giving $|F|^2>0.9$ up to $\theta\simeq90^\circ$
for $E_\nu=10$ MeV [Fig.~1].

\subsection{Effective Differential Cross Section for incoherent nucleons}
The effective cross section $d\Sigma/d\Omega$ for incoherent nucleons 
can be obtained by Eq. (5) putting $N=$1, $Z=$0 for neutrons and 
$N=$0, $Z=$1 for protons as 
\begin{align}
\frac{d\Sigma_{incoh}}{d\Omega}=
& \frac{G_F^2E_\nu^2}{4\pi^2}(1+\cos\theta)|F(q)|^2 (neutron) \notag \\
& (5.78\times10^{-3})\times \frac{G_F^2E_\nu^2}{4\pi^2}(1+\cos\theta)|F(q)|^2 (proton).
\end{align}

\subsection{Effective Differential Cross Section for coherent nuclei}
It is well know that the coherent scattering cross sections in the small-angle 
X-ray and neutron scatterings are enhanced by a factor of $N_{coh}^{2}$ 
where $N_{coh}$ stands for the number of the target nuclei coherently 
contributing to the scattering \citep{Lovesey1984}. 
Similarly, the effective cross section per unit volume $d\Sigma/d\Omega$ for 
neutrino-elastic scattering of many coherent nuclei can be written as
\begin{align}
\frac{d\Sigma}{d\Omega}
&= \frac{1}{V_{\rm eff}(q)}(n_tV_{\rm eff}(q))^{2}e^{-2W_{\rm bulk}}
\frac{d\sigma_{\rm single}}{d\Omega} \notag \\
&= n_t^2V_{\rm eff}(q)e^{-2W_{\rm bulk}}
\frac{d\sigma_{\rm single}}{d\Omega},
\end{align}
where $n_t$ is the number density of nuclei,
and $q$ is the momentum transfer,
\begin{equation}
q(\theta)=\frac{2E_\nu}{c}\sin(\theta/2).
\end{equation}
$d\sigma_{\rm single}/d\Omega$ stands for the differential CE$\nu$NS 
cross section for single nucleus introduced in the previous subsection. 

The effective coherence volume
\begin{equation}
V_{\rm eff}(q)=
\min[(2\pi/q)^3,(4\pi/3)(\xi a)^3]
\end{equation}
is bounded by the phenomenological correlation length $\xi a$,
with $a$ being the ion-sphere radius ($10^{-11}$–$10^{-10}$ cm in the post-shock region).
This cap prevents the $(2\pi/q)^3$ term from diverging at very small $q$;
thus $\xi$ acts as a low-$q$ cutoff parameter determined by the plasma coupling strength.

The factor $e^{-2W_{\rm bulk}}$ represents the Debye–Waller factor which accounts for 
suppression of coherence due to thermal fluctuation of the scatterers, given by  
\begin{equation}
e^{-2W_{\rm bulk}}=\exp[-q^2\langle(\Delta R)^2\rangle],
\end{equation}
where the mean-square relative displacement $\langle(\Delta R)^2\rangle$
is derived from the plasma frequency,
\begin{equation}
\omega_p^2=\frac{4\pi Z^2e^2n_t}{M},\qquad
\langle(\Delta R)^2\rangle=\frac{k_{\rm B}T}{M\omega_p^2}.
\end{equation}
For post-shock conditions ($T=2$–3 MeV, $\rho=10^{10}$–$10^{11}$ g cm$^{-3}$),
one obtains $\sqrt{\langle(\Delta R)^2\rangle} \approx 2.6\times10^{-12}$ cm = 26 fm and
$2W_{\rm bulk} \sim 7\times10^{-4}\ll1$, so multi-nuclear coherence is essentially preserved.

The number of coherent scatterers within a volume $V_{\rm eff}$ is
\begin{equation}
N_{\rm bulk}(q)\simeq n_tV_{\rm eff}(q)e^{-W_{\rm bulk}} ,
\end{equation}
and hence the volume cross section gains an enhancement factor $N_{\rm bulk}^2$.
For typical post-shock parameters
($\rho=10^{10}$ g cm$^{-3}$, $A=40$–100, $X_A\simeq0.5$),
$n_t\simeq3\times10^{31}$ cm$^{-3}$ and
$N_{\rm bulk}\sim10^3$–$10^4$ at forward angles,
yielding an enhancement $N_{\rm bulk}^2\sim10^6$ over incoherent scattering.
After solid-angle averaging, the gain remains about two orders of magnitude.

The above effective cross sections define the microscopic kernel $I(\theta)$ for 
the energy transfer process.
To verify internal consistency, we perform a simple numerical sanity check for 
a single-nucleus CE$\nu$NS case.
For $E_\nu = 10$ MeV, $\theta = 90^\circ$, $\rho = 10^{10}$ g cm$^{-3}$, and $A = 40$,
the standard differential cross section $d\sigma/d\Omega \simeq 6\times10^{-41}$ cm$^2$/sr
and the corresponding recoil energy $E_r\simeq2.5$ keV. Number density of nucleus is 
$n_{t} \simeq 1.5\times10^{31}$ cm$^{-3}$ and the effective coherent volume is $V_{\rm eff}=1/n_t$
for the single-nucleus limit (each nucleus occupying its own effective volume).
This yields $I(\theta)\simeq2\times10^{-3}$ keV per degree for $L=100$ km according to Eq. (2),
in excellent agreement with the dashed gray curve in Fig. 1.
This confirms that our scattering kernel reproduces the canonical CE$\nu$NS opacity
and that the normalization of the angular energy-transfer kernel is correct.
\begin{figure}[t]
\centering
\includegraphics[width=\linewidth]{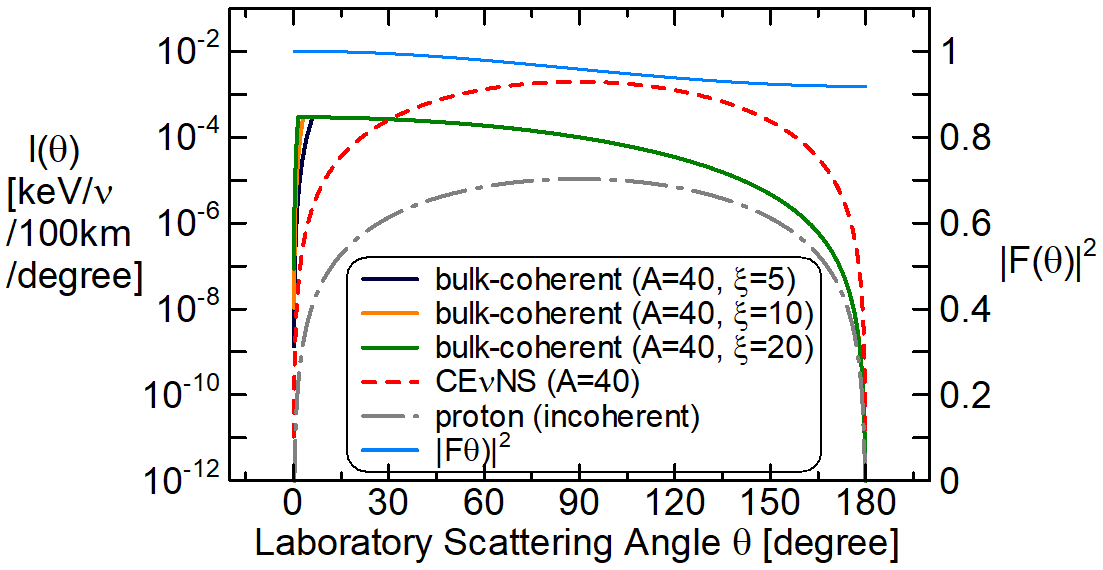}
\caption{Angular energy-transfer distribution
$I(\theta)$ for inter-nuclear coherence
($\xi=5$ (dark blue), 10 (orange), 20 (green)),
the incoherent limit (CE$\nu$NS, $\xi=1$, dashed red),
and incoherent proton scattering (dashed-dotted gray).
The nuclear form factor $|F(q)|^2$ is shown by the purple curve.
$E_\nu=10$ MeV and $A=40$ are used as representative parameters.}
\end{figure}

\section{Energetic impact of coherent scattering}

The volumetric heating rate produced by CE$\nu$NS can be estimated
by integrating the angular kernel $I(\theta)$ over the entire solid angle.
This converts the microscopic scattering rate (Fig.~1)
into a macroscopic energy-transfer rate shown in Fig.~2.

Figure 2 demonstrates that LET, integration of $I(\theta)$ across a $\sim$100 km
gain region yields several keV per incident neutrino.
The inter-nuclear coherent component becomes significant mainly below
$E_\nu\lesssim10$ MeV, where the long-wavelength correlation enhances
forward scattering.
At higher energies ($E_\nu\gtrsim10$ MeV), the dominant contribution arises
from single-nucleus CE$\nu$NS, giving $1.5$–$20$ keV per neutrino,
corresponding to roughly $0.015$–$0.2$\%  of the incident energy or 
equivalently (4.5-60)$\times$10$^{49}$ erg —covering the scale required 
to offset the typical $2$–$3\times10^{49}$ erg deficit in multidimensional 
CCSN models \citep{Sumiyoshi2005}.

For comparison, scattering on free protons produces negligible heating:
long-range correlations are suppressed by charge neutrality,
yielding $N_{\rm bulk}\approx1$ and
LET$\lesssim0.1$ keV for typical post-shock conditions, which is 
roughly two orders of magnitude smaller than those from CE$\nu$NS.  
This quantitative contrast is consistent with the finding of 
Thompson et al.~\citep{Thompson2003}, in which the inclusion of nucleon 
recoil produced no appreciable change in the explosion dynamics.  
Our results confirm that such incoherent recoils are intrinsically 
inefficient for energy deposition.

\begin{figure}[t]
\centering
\includegraphics[width=\linewidth]{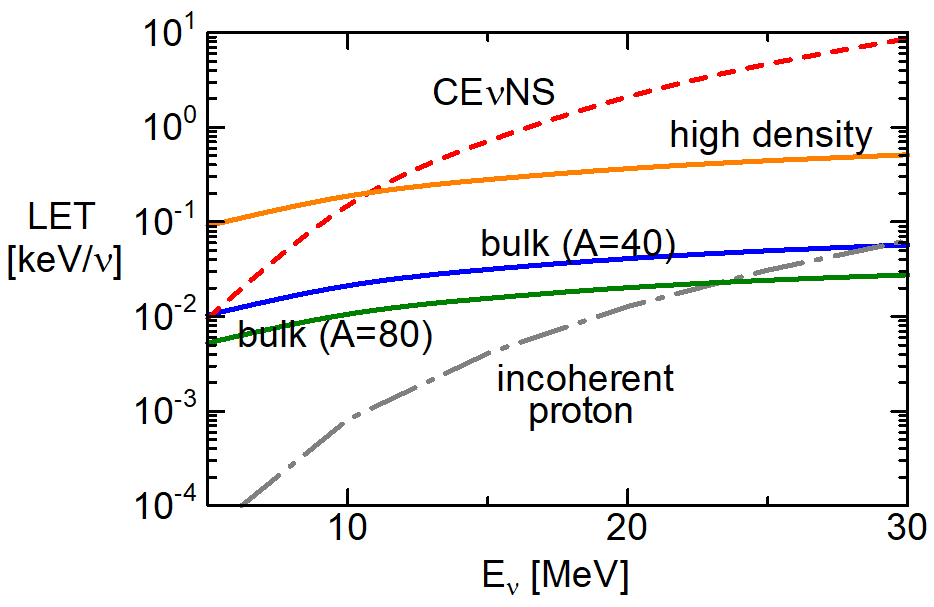}
\caption{Linear energy transfer (LET) as a function of neutrino energy.
Dark blue: baseline ($\xi=10$, $A=40$, $\rho=1\times10^{10}$ g cm$^{-3}$);
orange: heavy-nucleus rich ($\xi=10$, $A=80$);
green: higher-density case ($\xi=10$, $A=80$, $\rho=3\times10^{10}$ g cm$^{-3}$);
dashed red: single-nucleus CE$\nu$NS ($\xi=1$, $A=40$);
and dashed-dotted gray: incoherent proton scattering.
The CE$\nu$NS contribution reaches the magnitude needed for
shock revival.}
\end{figure}

\section{Summary}

We have reexamined neutrino–matter coupling in the post-shock region of core-collapse supernovae, 
emphasizing the role of nuclear recoil in coherent elastic neutrino–nucleus scattering (CE$\nu$NS).
Our calculations separate the microscopic scattering rate (Section 2) from the corresponding 
heating rate (Section 3), demonstrating quantitative consistency with standard CE$\nu$NS opacities.

We find that incoherent proton recoils are negligible, while coherent nuclear scattering can 
deliver a substantial linear energy transfer (LET) under typical post-shock conditions—\\
occasionally approaching the $\sim10^{49}$ erg energy increment required for shock revival.
Above $E_\nu\sim10$ MeV, CE$\nu$NS dominates the recoil-mediated energy transfer, because its 
bulk-\\
coherent enhancement scales as $q^{-3}$ and remains effective in the forward-scattering regime.
This demonstrates that even small per-event recoils, when integrated over the semi-transparent \\
$\sim$100 km region, can provide a dynamically relevant source of mechanical heating.

Historically, coherent weak neutral-current scattering was formulated by Freedman 
\citep{Freedman1974PRD}, and its potential importance in CCSN environments was pointed out 
by Sato \citep{KSato1975}.
Earlier transport analyses (e.g., Gershtein et al.\citep{Gershtein1975}) considered \\
thermonuclear-ignition scenarios in degenerate C/O cores, where the required energy scales 
differ by many orders of magnitude, so the mechanical recoil effect discussed here was 
naturally overlooked.

The smallness of nucleon recoils explains why previous \\
"recoil-included" 
simulations (e.g., \citep{Thompson2003,Burrows2018}) found little 
hydrodynamic effect: the dominant momentum-exchange channel\\
—coherent 
elastic $\nu$–A scattering—was still not taken into account. 
Only when recoil kinematics are restored in CE$\nu$NS does the accumulated 
energy transfer approach the canonical $(2–3)\times10^{49}$ erg scale 
required to revive the stalled shock.  
This result implies that the long-standing heating deficit arises not from 
an incorrect thermal budget, but from the historical omission of nuclear-scale 
recoil in coherent scattering. It should be noted that the inclusion of nuclear recoil 
constitutes a minimal yet physically grounded correction capable of bridging 
the gap between simulated and observed explosion energetics.

In this paper we have focused on the nuclear and proton components.
A fully consistent treatment using the dynamic structure factor $S(q,\omega)$, incorporating 
electrons, free neutrons, and realistic composition profiles from modern equations of state, 
is in progress and will be reported elsewhere. 
Such an approach will clarify the full impact of coherent recoil heating on the revival of 
the supernova shock.
Finally, because the CEvNS cross section depends strongly 
on the target nucleus, a realistic estimate of the neutrino energy deposition requires taking 
into account the isotopic mixture in the shocked matter. Work in this direction is in progress.

\section*{Acknowledgments}

The author is grateful to Profs.\ T.\ Kajino, K.\ Sumiyoshi, H.\ Togashi, and M.-K.\ Cheon
for fruitful discussions that inspired this study.
He also thanks Profs.\ S.\ Nagataki, T.\ Takiwaki, and S.\ Fujimoto
for clarifying the current treatment of recoil terms in state-of-the-art CCSN simulations.
Further thanks go to Profs.\ H.M.\ Shimizu and A.R.\ Young for valuable insights on
coherent neutron scattering, and to Prof.\ K.\ Ogata for comments on medium effects
in quantum scattering processes.

The author acknowledges helpful interactions with AI-based language models
(ChatGPT, OpenAI) for clarifying forward-scattering physics and improving
the manuscript presentation.
This work was partly supported by JSPS KAKENHI Grants 23K22502 and 22H01231.

\bibliographystyle{elsarticle-num}
\bibliography{plb_refs_revised}

\end{document}